\documentclass[aps, pre, twocolumn, superscriptaddress, reprint]{revtex4-1}

\usepackage{graphicx}
\usepackage{dcolumn}
\usepackage{bm}
\usepackage{amsmath, amssymb}

\usepackage{amsfonts, amsmath, bm, graphicx}
\usepackage{siunitx}
\usepackage{times}
\usepackage{lipsum}

\newcommand{\figref}[2]{\hyperref[#1]{\ref{#1}(#2)}}

\usepackage{lipsum}
\usepackage{changes}

\usepackage[colorlinks=true,allcolors=blue]{hyperref}
\usepackage{footmisc}

\usepackage{upgreek}

\usepackage{soul}

\begin{document}

{
\makeatletter
\def\frontmatter@thefootnote{%
 \altaffilletter@sw{\@fnsymbol}{\@fnsymbol}{\csname c@\@mpfn\endcsname}%
}%

\makeatother
\title{Nonlocal stimulation of three-magnon splitting in a magnetic vortex}

\author{L. K\"orber}\email{l.koerber@hzdr.de}
\affiliation{Helmholtz-Zentrum Dresden - Rossendorf, Institut f\"ur Ionenstrahlphysik und Materialforschung, D-01328 Dresden, Germany}
\affiliation{Fakult\"at Physik, Technische Universit\"at Dresden, D-01062 Dresden, Germany}

\author{K. Schultheiss}\email{k.schultheiss@hzdr.de}
\affiliation{Helmholtz-Zentrum Dresden - Rossendorf, Institut f\"ur Ionenstrahlphysik und Materialforschung, D-01328 Dresden, Germany}

\author{T. Hula}
\affiliation{Helmholtz-Zentrum Dresden - Rossendorf, Institut f\"ur Ionenstrahlphysik und Materialforschung, D-01328 Dresden, Germany}
\affiliation{Institut für Physik, Technische Universit\"at Chemnitz, Germany}

\author{R. Verba}
\affiliation{Institute of Magnetism, Kyiv 03142, Ukraine}

\author{J. Fassbender}
\affiliation{Helmholtz-Zentrum Dresden - Rossendorf, Institut f\"ur Ionenstrahlphysik und Materialforschung, D-01328 Dresden, Germany}
\affiliation{Fakult\"at Physik, Technische Universit\"at Dresden, D-01062 Dresden, Germany}

\author{A. K\'{a}kay}
\affiliation{Helmholtz-Zentrum Dresden - Rossendorf, Institut f\"ur Ionenstrahlphysik und Materialforschung, D-01328 Dresden, Germany}

\author{H. Schultheiss}
\affiliation{Helmholtz-Zentrum Dresden - Rossendorf, Institut f\"ur Ionenstrahlphysik und Materialforschung, D-01328 Dresden, Germany}
\affiliation{Fakult\"at Physik, Technische Universit\"at Dresden, D-01062 Dresden, Germany}

\date{\today}

\begin{abstract}
We present a combined numerical, theoretical and experimental study on stimulated three-magnon splitting in a magnetic disk in the vortex state. Our micromagnetic simulations and Brillouin-light-scattering results confirm that three-magnon splitting can be triggered even below threshold by exciting one of the secondary modes by magnons propagating in a waveguide next to the disk. The  experiments show that stimulation is possible over an extended range of excitation powers and a wide range of frequencies around the eigenfrequencies of the secondary modes. Rate-equation calculations predict an instantaneous response to stimulation and the possibility to prematurely trigger three-magnon splitting even above threshold in a sustainable manner. These predictions are confirmed experimentally using time-resolved Brillouin-light-scattering measurements and are in a good qualitative agreement with the theoretical results. We believe that the controllable mechanism of stimulated three-magnon splitting could provide a possibility to utilize magnon-based nonlinear networks as hardware for neuromorphic computing.
\end{abstract}

\maketitle

With the first experiments on ferromagnetic resonance at large excitation powers by Bloembergen, Damon and Wang \cite{Bloembergen1952,Damon1953,Bloembergen1954}, nonlinear effects in magnetization dynamics have often been regarded as parasitic, leading to additional losses, frequency shift or linewidth broadening of the ferromagnetic resonance and of the magnon modes therein. On the other hand, this nonlinearity leads to many interesting phenomena such as multi-magnon scattering, parametric pumping, formation of solitons or phase locking \cite{l2012wave, gurevich1996magnetization}. Some of these effects have analoga in other fields, like nonlinear optics \cite{bloembergen1991nonlinear, hasegawa2012physics}.

In recent years, nonlinear systems have been given increasing attention, for example, as candidates for hardware implementations of neuromorphic computing \cite{woods2012photonic,tanaka2019recent, markovic2019reservoir,torrejon2017neuromorphic}. The magnons (spin waves) in ferromagnets are strongly coupled and, above a certain threshold amplitude, undergo spontaneous scattering processes with each other, leading to information transport in wave-vector space. Recently, we provided the direct experimental evidence for three-magnon splitting (3MS), {\it i.e.}, the splitting of one primary magnon in two secondary magnons, in magnetic disks in vortex state \cite{Schultheiss19}. Moreover, Zhang and coworkers predicted numerically the stimulation of such processes in domain walls \cite{Zhang:2018fp}. However, up to now, there was no experimental demonstration of a nonlocal, active stimulation of magnon splitting in quantized systems, which allows not only to control the time, when the splitting sets in, but also to enforce the scattering into specific states.

\begin{figure}[t]
    \centering
    \includegraphics{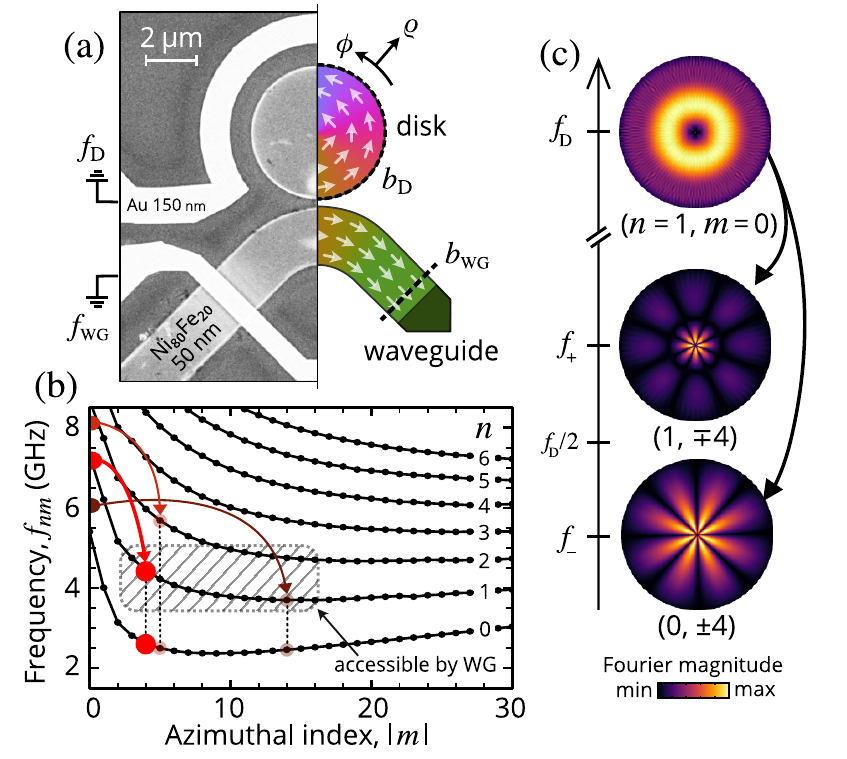}
    \caption{{(a)} Experimental realization (left) and  simulation model (right) of the vortex-state permalloy disk and of the longitudinally-magnetized \SI{2}{\micro\meter} wide waveguide. The experimental image has been acquired using scanning electron microscopy. An $\Omega$-shaped antenna is used to excite the disk with an RF field $b_\mathrm{D}$ at frequency $f_\mathrm{D}$. Additionally, a strip-line antenna is used to excite magnons in the waveguide with an RF field $b_\mathrm{WG}$ at frequency $f_\mathrm{WG}$. {(b)} Dispersion of the vortex disk calculated using micromagnetic simulations. Exemplary scattering channels for three different excitation frequencies are marked on the dispersion. The hatched rectangular region represents an approximate frequency and wave-vector interval in which the waveguide can efficiently excite magnons in the disk. {(c)} Numerically obtained spatial profiles of the magnons taking part in a 3MS channel: the directly excited radial mode with frequency $f_\mathrm{D}$ and the secondary azimuthal modes with $f_{+}$ and $f_{-}$.} 
    \label{fig:fig1}
\end{figure}

In this Letter, we use rate-equation calculations, micromagnetic simulations and Brillouin-light scattering microscopy ($\upmu$BLS) to predict and verify that 3MS can be triggered even below threshold by exciting one of the secondary modes via the dynamic magnetic fields of magnons propagating in a waveguide next to the disk (Fig.~\figref{fig:fig1}{a}). The experiments show that stimulation is possible at low powers and in a wide frequency range around the frequencies of the secondary modes. Furthermore, we demonstrate the instantaneous response to stimulation and the possibility  to prematurely trigger 3MS above threshold.

Magnon eigenmodes in vortex-state magnetic disks have been examined in numerous experimental \cite{Buess04, Buess05, Schultheiss19, Slobodianiuk19}, theoretical \cite{Slobodianiuk19, Guslienko08, Ivanov05, Verba2016, Guslienko02JAP, Galkin06, Sheka2004PRB, Ivanov98, Ivanov02a} as well as numerical works \cite{Gang13, Neudecker06, Taurel16}. Their confinement leads to a discrete spectrum which is categorized by a radial index $n \geq 0 $, counting the nodes in the radial ($\varrho$) direction, and an azimuthal index $m \in \mathbb{Z}$, counting the periods in the azimuthal $\phi$-direction (see coordinate system in Fig.~\figref{fig:fig1}{a}). These modes appear in degenerate duplets. For large enough disks, two modes with the same $n$ but opposite $m$ share the same frequency, with the exception of $m=\pm 1$ for which hybridization with the gyrotropic mode of the vortex appears \cite{Guslienko08, Ivanov05,Hoffmann07,Buess04}. Fig.~\figref{fig:fig1}{b} shows the spectrum of the vortex magnons obtained by micromagnetic simulations for a permalloy (Ni$_{80}$Fe$_{20}$) disk of \SI{50}{\nano\meter} thickness and \SI{5.1}{\micro\meter} diameter (see Supplemental Material). The slope of the dispersion is negative in azimuthal direction (increasing $|m|$) and positive in radial direction (increasing $n$), which is characteristic for dipolar-dominated spin-wave propagation parallel and perpendicular to the equilibrium magnetization in thin magnetic films \cite{Buess04, Buess05, Kalinikos86}.

A radial mode ($m=0$), that is excited by an azimuthally symmetric out-of-plane microwave (RF) field at frequency $f_\mathrm{D}$ and with a large enough power, may decay into two secondary modes via 3MS \cite{Schultheiss19}, as shown using micromagnetic simulations in Fig.~\figref{fig:fig1}{c} for an excitation frequency of $f_\mathrm{D}=\SI{7.2}{\giga\hertz}$. Experimentally, such an RF field can be achieved by an $\Omega$-shaped microwave antenna around the magnetic disk. The 3MS in magnetic vortices obeys certain selection rules \cite{Schultheiss19}: Because of the cylindrical symmetry, the angular momentum in $\phi$-direction must be conserved, resulting in secondary modes of opposite azimuthal index $\pm m$. Moreover, the two secondary modes must have different radial indices which leads to a pronounced splitting in frequency $2\Delta f$ between them. Finally, energy conservation demands for the frequencies of the secondary modes $f_{\pm}=f_\mathrm{D}/2 \pm \Delta f$. Simultaneously, the mirrored process with exchanged signs of the azimuthal indices is equally probable. We associate these two equivalent processes with one scattering channel and speak of secondary duplets instead of secondary modes. This symmetry leads to standing waves which can be observed, \textit{e.g.}, by $\upmu$BLS \cite{Schultheiss19, Sebastian15}. Depending on the excitation frequency of the radial mode $f_\mathrm{D}$, a variety of scattering channels can be observed, some of which are marked on the dispersion in Fig.~\figref{fig:fig1}{b}.

For all channels, the frequency split $2\Delta f$ between the secondary duplets can be exploited to access them individually using an additional RF field at the respective frequency $f_+$ or $f_-$. It is well known that 3MS leads to a considerable feedback on the directly excited mode at $f_\mathrm{D}$. This interconnection together with the fact that the secondary duplets could be excited individually raised the question whether stimulated splitting could be realized even below threshold. In other words, what happens if we excite one of the secondary duplets directly (\textit{e.g.}, at $f_+$), taken that the radial mode at $f_\mathrm{D}$ is excited below threshold. A direct evidence for stimulated 3MS would be an instant response at $f_-$.

To answer this question, we have performed micromagnetic simulations using a custom version of MuMax$^3$ \cite{Vansteenkiste14,MuMax2} and verified these numerical results by $\upmu$BLS. In order to resonantly excite one of the secondary duplets (with $\vert m\vert>0$), it is not possible to use an $\Omega$-shaped antenna. The reason is that these modes possess no net magnetic moment and, therefore, cannot couple to the azimuthally symmetric RF field produced by such an antenna. Instead, we use an adjacent, \SI{2}{\micro\meter} wide waveguide (WG) of the same material and thickness as the disk. In this waveguide, we inject propagating magnons at frequency $f_\mathrm{WG}$ that couple to the azimuthal modes within the disk. A curved waveguide is used to allow for a better surrounding of the disk by the $\Omega$-shaped antenna. To maximize their intensity within the waveguide, magnons are excited at both ends resulting in standing waves in the vicinity of the disk. The experimental and numerical sample design is shown in Fig.~\figref{fig:fig1}{a}. An approximate frequency and wave-vector regime, in which azimuthal modes in the disk can be excited by the waveguide, is marked in the spectrum in Fig.~\figref{fig:fig1}{b}.

Without loss of generality, stimulated 3MS is shown for the channel at $f_\mathrm{D}=\SI{7.20}{\giga\hertz}$, introduced in Fig.~{\figref{fig:fig1}{c}}, with the secondary duplets at $f_+=\SI{4.46}{\giga\hertz}$ and $f_-=\SI{2.74}{\giga\hertz}$. Confirmation for other channels is found in the Supplemental Material.

\begin{figure}[b]
    \centering
    \includegraphics{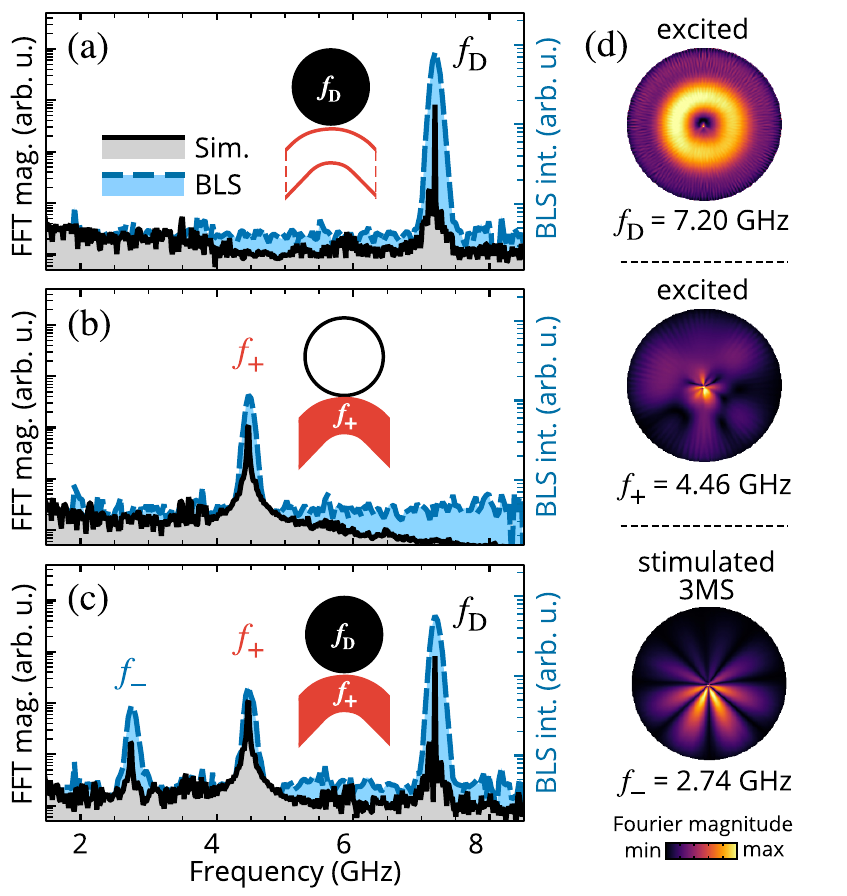}
    \caption{Spectra of the vortex disk obtained by micromagnetic simulations and $\upmu$BLS for three cases: {(a)} exciting only the disk below threshold at frequency $f_\mathrm{D}$, {(b)} exciting only the waveguide at frequency $f_+$ and {(c)} both combined, showing an additional signal at $f_-$. {(d)} mode profiles obtained from the simulation in (c).}
    \label{fig:gallery-stimulation}
\end{figure}

As a first sanity check, we excite (using the $\Omega$-shaped antenna in the experiments) the radial mode $(1,0)$ in the disk at frequency $f_\mathrm{D}$ with a low power to verify that the direct excitation is below threshold and no secondary duplets are observed \footnote{In simulations, the RF field was set to about 75\,\% of the threshold field. In experiments, an excitation at about 50\,\% of the threshold power was used.}. The corresponding numerical and experimental spectra are shown in Fig.~{\figref{fig:gallery-stimulation}{a}}.
In the simulations, the spectrum of the disk was obtained by performing a Fourier transform at each cell in the disk whereas the experimental spectrum was obtained by measuring and integrating the $\upmu$BLS spectra at 14 different points on the disk.

Next, to verify that magnons in the waveguide can couple to the azimuthal modes in the disk, we pump the waveguide at the frequency of the higher duplet $f_\mathrm{WG} = f_+$. By again measuring the spectral response in the disk (Fig.~\figref{fig:gallery-stimulation}{b}), we observe a signal at $f_+$, confirming a successful coupling. 
Here, no RF field at the {$\Omega$}-shaped antenna was applied.

Finally, we combine both schemes by exciting at the frequency of the  radial mode $f_\mathrm{D}=\SI{7.20}{\giga\hertz}$ in the disk and at 
$f_+=\SI{4.46}{\giga\hertz}$ in the waveguide. As seen in Fig.~\figref{fig:gallery-stimulation}{c}, we observe an additional spectral contribution at the exact frequency of the lower duplet $f_-=\SI{2.74}{\giga\hertz}$. By obtaining the mode profiles at $f_-$ and $f_+$ from micromagnetic simulations (Fig.~\figref{fig:gallery-stimulation}{d}) -- especially in comparison with the profiles of spontaneous 3MS at the same excitation frequency $f_\mathrm{D}$ in Fig.~\figref{fig:fig1}{c} -- we can confirm that the signals truly belong to azimuthal modes. As expected, these modes show different radial indices $n$. The profile of the mode excited at $f_+$ by the magnons in the waveguide is distorted. As the dipolar field of the waveguide generates a wide spectrum of $|m|$-components at the site of the disk, we attribute this to an excitation of multiple degenerate duplets at that frequency. However, the profile of the solely parametrically excited mode $f_-$ almost perfectly resembles the profile of the lower duplet produced by spontaneous 3MS. Thus, we conclude that 3MS has been successfully stimulated below threshold.

In the remaining part of this Letter, we want to explore three aspects of stimulated 3MS, namely the bandwidth of stimulation, the dependence on the excitation power within the waveguide as well as the temporal evolution of the magnons subject to stimulated splitting.


\begin{figure}[h]
    \centering
    \includegraphics[width=8.6cm]{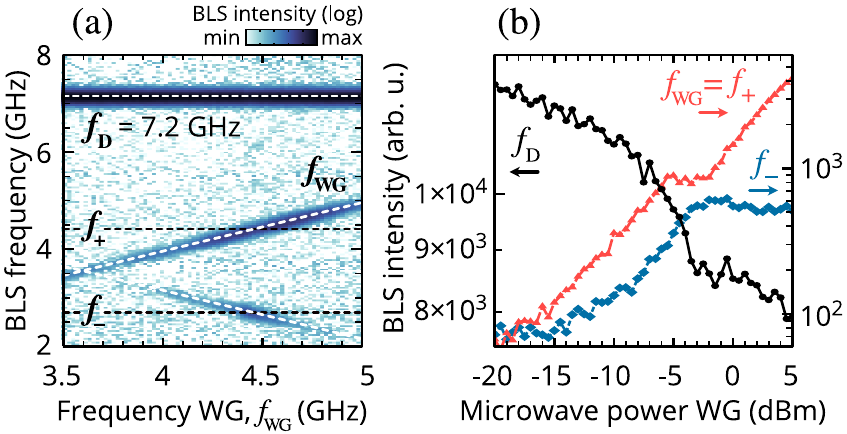}
        \caption{(a) BLS spectra of the disk when excited below threshold at frequency $f_\mathrm{D}$ and the excitation frequency in the waveguide $f_\mathrm{WG}$ is varied. The response at $f_-$ corresponds to the lower duplet, as a result of stimulated 3MS. In (b), the BLS response is extracted by integrating over a frequency interval around the respective frequencies when the disk is excited at $f_\mathrm{D}=\SI{7.2}{\giga\hertz}$ below threshold and the excitation power in the waveguide is varied at constant frequency $f_\mathrm{WG}=f_+=\SI{4.46}{\giga\hertz}$. Arrows indicate the corresponding logarithmic intensity axis.}
    \label{fig:fig3}
\end{figure}
To address the bandwidth, 
we excite the disk below threshold at {$f_\mathrm{D}$} and vary the excitation frequency {$f_\mathrm{WG}$} in the waveguide.
Fig.~\figref{fig:fig3}{a} shows the $f_\mathrm{WG}$-dependent spectral response of the disk obtained by $\upmu$BLS. The dashed diagonal line at $f_\mathrm{WG}$ is accompanied with an anti-diagonal response over a wide frequency range of about $\SI{0.6}{\giga\hertz}$ around the frequencies of the duplets. This confirms that stimulated 3MS is possible even when $f_\mathrm{WG}$ does not exactly match the frequency of one of the secondary duplets \cite{Slobodianiuk19, 6595654}.

Next, we address the influence of the power, that the waveguide is excited with. In the experiment, we fix $f_\mathrm{WG}=f_+$ and vary the microwave power applied to the antenna on the waveguide. The BLS intensities of the modes in the disk were extracted by integrating over a certain linewidth around the respective frequencies. The results summarized in Fig.~\figref{fig:fig3}{b} confirm that stimulated splitting is possible even for low excitation powers in the waveguide. With increasing power of $f_\mathrm{WG}$, the measured signal at $f_-$ increases whereas the one at $f_\mathrm{D}$ decreases. This demonstrates the energetic interconnection between the different modes and is an indirect evidence that a splitting process is taking place.
We observe a saturation in the power of the stimulated splitting above which the intensity of the parametrically excited duplet at $f_-$ does not increase further. However, the resonantly excited duplet at $f_\mathrm{WG}= f_+$ still increases. As seen from the evolution of $f_\mathrm{D}$ in Fig.~\figref{fig:fig3}{b}, above this power, the energy flux supplied by the directly excited radial mode is nearly exhausted. Naturally, the position of the saturation depends on the excitation power of this radial mode.
\begin{figure*}
    \centering
    \includegraphics[width=17cm]{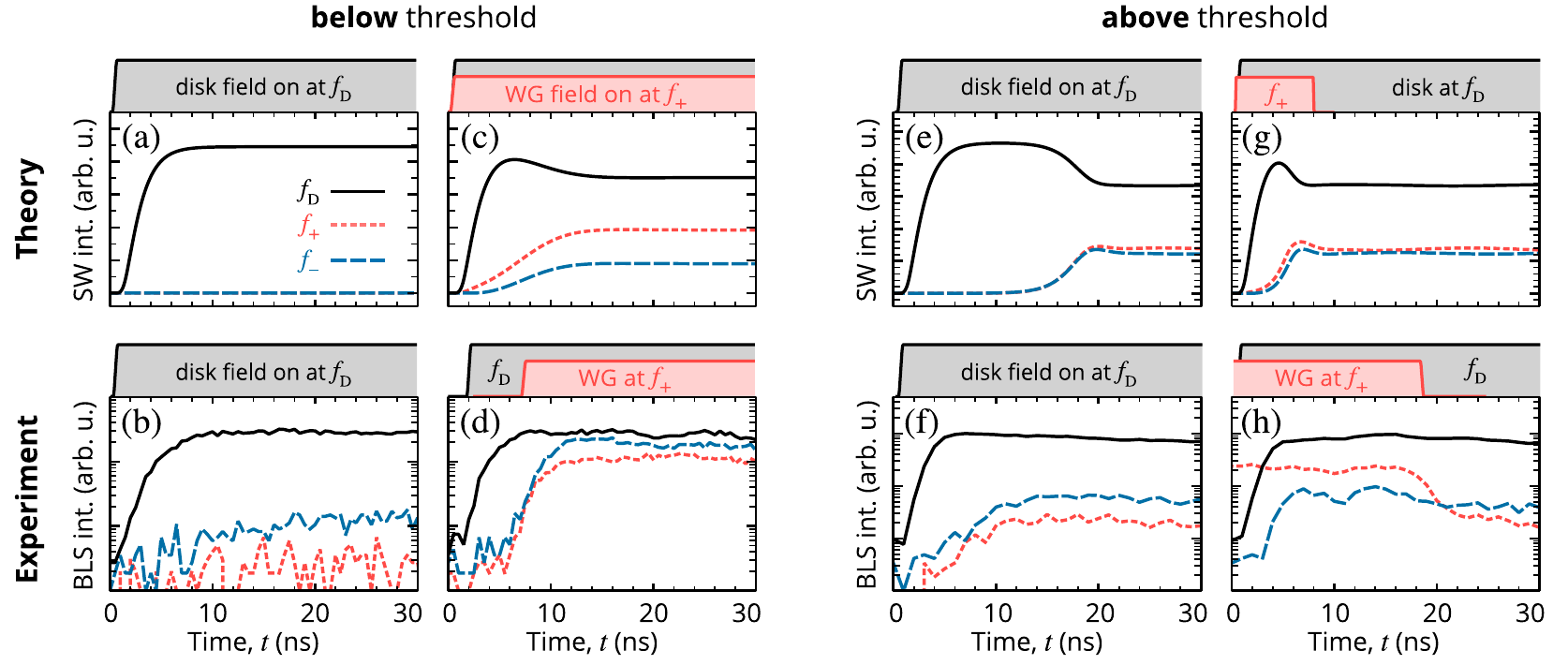}
        \caption{
        Temporal evolution of the spin-wave intensities obtained by nonlinear spin-wave theory and TR-$\upmu$BLS, respectively. The time frames when the RF fields are turned on are marked by the boxes on top of the axes. In (a,\,b), only the radial mode at $f_\mathrm{D}$ is excited below threshold. In (c,\,d), an additional RF field is applied at $f_+$ in the waveguide, showing a successful stimulation of the mode at $f_-$. (e,\,f) show spontaneous 3MS above threshold, whereas (g,\,h) show a shortening of the parametric time delay by applying an RF pulse at $f_+$ in the waveguide. The ratio between the intensity levels of different modes in the experiments deviate from the theoretical calculation, likely due to the spatial dependence of the duplets which form standing waves. Since TR-$\upmu$BLS measurements are time-consuming they were only performed at one position on the disk. Different scales were used for theory and experiments to highlight the feedback on the directly excited mode from theory, but also to show the small signals of TR-$\upmu$BLS.
        }
    \label{fig:fig4}
\end{figure*}
Finally, we focus on how stimulation provides means to control the time scale of 3MS. Parametric phenomena of this nature are known to exhibit a power-dependent delay. It takes a certain time before the directly excited mode reaches its threshold, before the secondary duplets start to grow, and even a longer time before these magnons reach their dynamic equilibrium. Since stimulated splitting is possible below threshold, the temporal evolution of the modes must change compared to spontaneous scattering. For this, we utilize rate equations derived from nonlinear spin-wave theory \cite{l2012wave, Krivosik10, livesey2015nonlinear}, combined with time-resolved $\upmu$BLS (TR-$\upmu$BLS). Both methods allow us to track the temporal evolution of magnons (see Supplemental Material for details). Only a qualitative comparison between theory and experiments is presented here because, first, the exact power arriving at the antenna is unknown in the experiment and, second, for the same reason, a synchronization of the microwave pulses at the sample is cumbersome. Moreover, in the theoretical model, an additional delay arising from the finite group velocity of the magnons excited in the waveguide is neglected.

Let us first focus on the case below threshold. In Fig.~\figref{fig:fig4}{a,\,b}, we show the reference case when only the radial mode at $f_\mathrm{D}$ is excited with a RF field, marked with a box on top of the panels. As soon as the duplet at $f_+$ is excited as well, the parametrically excited duplet at $f_-$ immediately follows (Fig.~\figref{fig:fig4}{c,\,d}). This means, that the response to stimulation is almost instant. Note that, here, we observe a feedback on the directly excited mode, \textit{i.e.}, a loss in its intensity due to the opening of the 3MS channel. This fast response can be used to shorten the power-dependent parametric time delay above threshold. If again only the radial mode is excited, now above threshold, we observe spontaneous 3MS. The secondary modes reach dynamic equilibrium at about \SI{20}{\nano\second} in the theoretical calculation (Fig.~\figref{fig:fig4}{e}) and at \SI{15}{\nano\second} in the experiment (Fig.~\figref{fig:fig4}{f}). Also here, the secondary modes appear much earlier if stimulated 3MS takes place (Fig.~\figref{fig:fig4}{g,\,h}). 
Additionally, in the theoretical calculation, the RF field in the waveguide at $f_+$ is only applied for a short duration at the beginning of the RF pulse at $f_\mathrm{D}$, \textit{i.e.}, we stimulate only for a short time. Note that this time is much smaller than the time needed for the secondary modes to reach dynamical equilibrium in the case of spontaneous 3MS in Fig.~\figref{fig:fig4}{e}. After the stimulation pulse at $f_+$ in Fig.~\figref{fig:fig4}{g} is turned off, spontaneous 3MS takes over and all modes relax slowly into their dynamic equilibrium. This illustrates that 3MS can be prematurely triggered above threshold even using only a short stimulation pulse.

In conclusion, we demonstrated that stimulated 3MS in a magnetic vortex can be achieved by coupling the dynamic fields of magnons propagating in an adjacent waveguide to the disk. Using micromagnetic simulations, nonlinear spin-wave theory and Brillouin-light-scattering microscopy, we have predicted and confirmed that three-magnon splitting can be triggered below threshold by exciting one of the secondary modes/duplets. The BLS experiments have shown that stimulation is possible even at low powers and in a wide bandwidth around the frequencies of the secondary modes/duplets. Finally, we showed the instantaneous response to stimulation and the possibility to even prematurely trigger 3MS above threshold in a sustainable manner. Our theoretical predictions 
are not only verified experimentally but are qualitatively in good agreement. We believe that stimulated 3MS provides a new possibility to harness the potential of nonlinear spin-wave dynamics. 
An ensemble of vortex disks coupled to a network of waveguides could be utilized in magnon-based nonlinear networks. Since the temporal evolution of the secondary modes resembles a neuron response function, such networks using stimulated 3MS could serve as a candidate for hardware 
neuromorphic computing.\\

Financial support by the Deutsche Forschungsgemeinschaft within programmes SCHU 2922/1-1 and KA 5069/1-1 is gratefully acknowledged. K.S. acknowledges funding within the Helmholtz Postdoc Programme. R.V. acknowledges support of NAS of Ukraine through project No. 23-04/01-2020. The samples were fabricated at the Nanofabrication Facilities (NanoFaRo) at the Institute of Ion Beam Research and Materials Research at HZDR. We thank B. Scheumann for the film deposition. L.K. and K.S. equally contributed to this work.

\bibliography{vortex}

\end{document}